\documentclass[aps,prl,twocolumn,superscriptaddress,showpacs,floatfix]{revtex4-1}

\usepackage{hyperref}
\usepackage{color}
\usepackage{graphicx}	
\usepackage[utf8]{inputenc} 
\graphicspath{%
  {./},%
}

\usepackage{amstext}
\usepackage{amsmath}
\usepackage{array}
\usepackage{enumitem} 
\usepackage{bm}
\usepackage{upgreek}

\usepackage{xspace}	

\begin{document}

\title{Fluid Dynamics Far From Local Equilibrium}

\author{Paul Romatschke} 
\affiliation{Department of Physics, University of Colorado, Boulder, Colorado 80309, USA}
\affiliation{Center for Theory of Quantum Matter, University of Colorado, Boulder, Colorado 80309, USA}
\date{\today}

\begin{abstract}
Fluid dynamics is traditionally thought to apply only to systems near local equilibrium. In this case, the effective theory of fluid dynamics can be constructed as a gradient series. Recent applications of resurgence suggest that this gradient series diverges, but can be Borel-resummed, giving rise to a hydrodynamic attractor solution which is well defined even for large gradients. Arbitrary initial data quickly approaches this attractor via non-hydrodynamic mode decay. This suggests the existence of a new theory of far-from-equilibrium fluid dynamics. In this work, the framework of fluid dynamics far from local equilibrium for conformal system is introduced, and the hydrodynamic attractor solutions for rBRSSS, kinetic theory in the relaxation time approximation, and strongly-coupled N=4 SYM are identified for a system undergoing Bjorken flow. 
  \end{abstract}

\maketitle

What is fluid dynamics and what is its regime of applicability? Over the centuries, different answers have been given to this question.

The textbook definition of the applicability of fluid dynamics is that the local mean free path should be much smaller than the system size. This criterion originates from the notion that fluid dynamics is the macroscopic limit of some underlying kinetic theory. In kinetic theory, the mean free path has the intuitive interpretation of the typical length a particle can travel before experiencing a collision. If that mean free path length is larger than the system size, particles will not experience collisions before leaving the system, thus invalidating a fluid dynamic description.

In relativistic fluid dynamics' modern formulation, the phenomenological ‘mean-free-path’ criterion is replaced by the requirement that gradients around some reference configuration (typically local equilibrium) are small when compared to system temperature. This gives rise to the notion of fluid dynamics as the effective theory of long-wavelength excitations, which can be expressed as a hydrodynamic gradient series. In this framework, the Navier-Stokes equations arise as the unique theory that is defined by the most general energy-momentum tensor that can be built out of hydrodynamic fields and first-order gradients thereof.

Thus, even in the modern framework, the requirement of small gradients seems to limit the applicability of fluid dynamics to the near-equilibrium regime.

However, there is mounting evidence that (first or second-order) fluid dynamics offers a correct quantitative description of systems which are not close to local equilibrium. For instance, a variety of numerical experiments indicate that fluid dynamics can match exact results even if the gradient corrections (normalized by temperature) are of order unity \cite{Chesler:2009cy,Heller:2011ju,Wu:2011yd,vanderSchee:2012qj,Casalderrey-Solana:2013aba,Kurkela:2015qoa,Keegan:2015avk}. Experimentally, ultrarelativistic collisions of protons exhibits the same flow features as much larger systems produced in heavy-ion collisions \cite{Aad:2015gqa,Khachatryan:2015lva}. Despite gradients in proton collisions being large, low-order hydrodynamics offers a quantitatively accurate description of experimental flow results \cite{Bozek:2010pb,Werner:2010ss,Weller:2017tsr}. This suggests that the mean-free path criterion 
for the applicability of fluid dynamics is possibly too strict and should be replaced by the ability to neglect the effect from  non-hydrodynamic modes \cite{Romatschke:2016hle}.

One may now wonder if the (unreasonable?) success of low-order fluid dynamics is perhaps caused by a particularly rapid convergence of the hydrodynamic gradient expansion. To test this idea, the gradient series coefficients were calculated for specific microscopic theories to very high order, cf. Refs.~\cite{Heller:2013fn,Heller:2015dha,Basar:2015ava,Buchel:2016cbj,Heller:2016rtz,Denicol:2016bjh,Florkowski:2016zsi}. Curiously, it was found that the gradient series not only is not rapidly convergent, but is actually a divergent series. However, it seems that this divergent series is Borel-summable in a generalized sense. In a groundbreaking article, Heller and Spali{\'n}ski demonstrated that the Borel-resummed gradient series leads to the presence of a unique hydrodynamic ‘attractor’ solution \cite{Heller:2015dha}. Arbitrary initial data in the underlying microscopic theory quickly evolves towards this unique attractor solution. In the limit of small gradients, the attractor reduces to the familiar low-order hydrodynamic gradient series solution.

Perhaps more interestingly, while the hydrodynamic gradient series diverges, 
the (non-analytic) hydrodynamic attractor solution is very well approximated by the low order hydrodynamic gradient series even for moderate gradient sizes, at least for the system studied in Ref.~\cite{Heller:2015dha}. This is typical for solutions which possess an asymptotic series expansion, cf. the case of perturbative QCD at high temperature \cite{Blaizot:2003iq}. As argued in Ref.~\cite{Romatschke:2016hle}, this observation naturally explains the success of low-order hydrodynamics in accurately describing out-of-equilibrium systems where gradients are of order unity.

In the present work, I will attempt to generalize the notion of fluid dynamics to systems with conformal symmetry far from local equilibrium,  where normalized gradients are not only of order unity, but large. This generalization requires the presence of a non-analytic hydrodynamic attractor solution far from equilibrium. Because arbitrary initial data will typically not fall onto this attractor solution, far-from-equilibrium fluid dynamics will not describe most far-from-equilibrium solutions the microscopic dynamics generates. In this sense, far-from-equilibrium fluid dynamics is no replacement for solving the exact microscopic dynamics for a specific initial condition. However, the reason why far-from-equilibrium evolution does not match the hydrodynamic attractor is that for arbitrary initial data, other, non-hydrodynamic modes are typically excited \cite{Heller:2013fn,Heller:2015dha,Brewer:2015ipa,Romatschke:2015gic,Grozdanov:2016vgg,Florkowski:2016zsi}. These non-hydrodynamic modes are specific to the microscopic theory under consideration, but have in common that they decay on time-scales short compared to the typical fluid dynamic time scale (e.g. the mean free path) as long as hydrodynamic modes exist \cite{Romatschke:2015gic,Grozdanov:2016vgg,Romatschke:2016hle}. Hence all solutions for arbitrary initial data will eventually merge with the hydrodynamic attractor after the non-hydrodynamic modes have decayed.


\paragraph{\bf Near Equilibrium Fluid Dynamics}

Let us consider a conformal quantum system in four dimensional Minkowski space-time and  assume that the expectation value of the energy-momentum tensor $\langle T^{\mu\nu}\rangle$ in equilibrium can be calculated. In this case, $\langle T^{\mu\nu}\rangle$ possesses a time-like eigenvector $u^\mu$, normalized to $u^\mu u_\mu=-1$,  with an associated eigenvalue $\epsilon$ (I will be  using the mostly plus convention for the metric tensor $g_{\mu\nu}$). It is then straightforward to show that $\langle T^{\mu\nu}\rangle$ can be decomposed as the ideal fluid energy-momentum tensor $T^{\mu\nu}_{(0)}=(\epsilon+P)u^\mu u^\nu+P g^{\mu\nu}$, where $P$ is the equilibrium pressure for this four-dimensional conformal theory. Since for a conformal system in flat space $T^\mu_\mu=0$, the equilibrium equation of state is $P(\epsilon)=\epsilon/3$.

Near-equilibrium corrections to the ideal fluid energy-momentum tensor can systematically be derived by considering all possible independent symmetric rank two tensors with one, two, three, etc. gradients consistent with conformal symmetry \cite{Baier:2007ix,Bhattacharyya:2008jc,Grozdanov:2015kqa}. The first-order correction is given by
\begin{equation}
\label{eq:NS}
T^{\mu\nu}_{(1)}=-\eta \sigma^{\mu\nu}\,,
\end{equation}
where $\eta$ is the shear viscosity coefficient and $\sigma^{\mu\nu}=2 \nabla_\perp^{(\mu} u^{\nu)}-\frac{2}{3}\Delta^{\mu\nu}\nabla_\lambda u^\lambda$ where $\nabla_\perp^\mu=\Delta^{\mu\nu}\nabla_\nu$ and $\Delta^{\mu\nu}=\left(g^{\mu\nu}+u^\mu u^\nu\right)$. As was the case in equilibrium, the energy density $\epsilon$ and flow vector $u^\mu$ can still be defined as the time-like eigenvalue and eigenvector of the microscopic energy-momentum tensor $\langle T^{\mu\nu}\rangle$ as long as a local rest frame exists \cite{Arnold:2014jva}. The conservation of $T^{\mu\nu}_{(0)}+T^{\mu\nu}_{(1)}$ constitutes a near-equilibrium fluid dynamic theory, since $T^{\mu\nu}_{(1)}$ should be a small correction to $T^{\mu\nu}_{(0)}$.
For higher order corrections, counting the number of independent structures built out of contractions of $\nabla_{\mu}$, one finds that there are at least $(n-2)!$ terms contributing to $T^{\mu\nu}_{(n)}$. Assuming the coefficients multiplying these terms do not decrease too fast, this implies that $T^{\mu\nu}_{(n)}$ grows as $n!$ for fixed gradient strength $\nabla_\mu$. Hence one can expect the hydrodynamic gradient series to diverge for any non-vanishing gradient strength.

\paragraph{\bf Borel Resummation}

\begin{figure*}[t]
  \includegraphics[width=\linewidth]{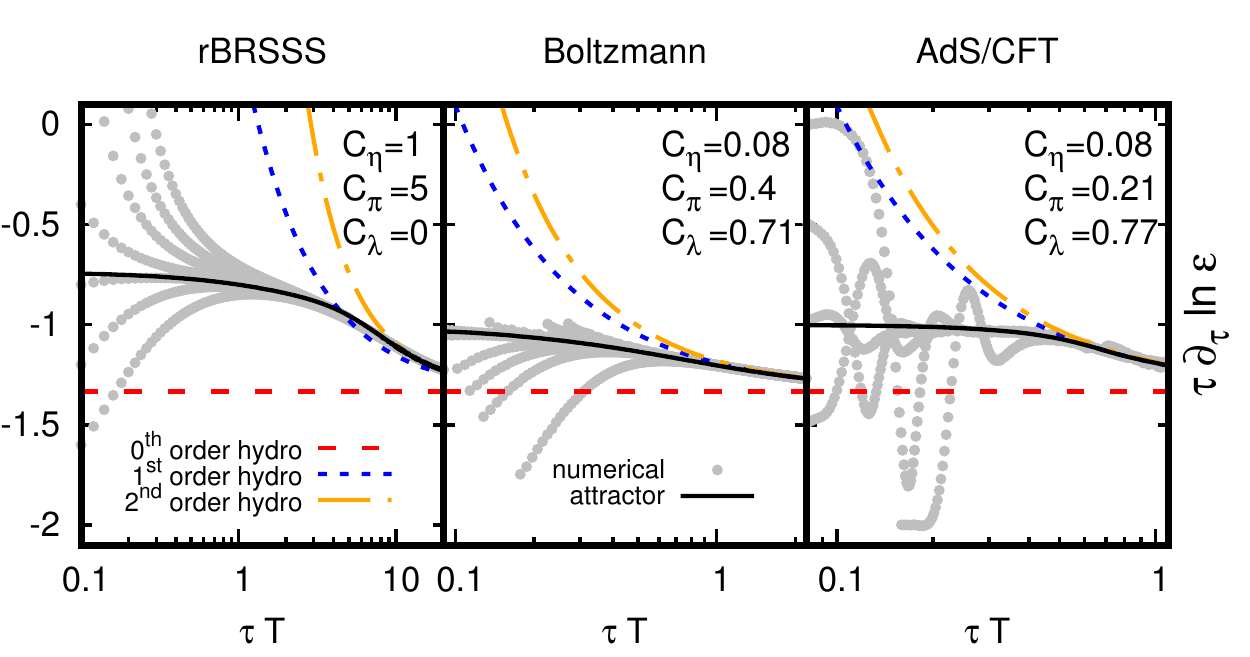}
  \caption{\label{fig:mis} Numerical results for energy density evolution as a function of inverse gradient strength $\tau T$ for conformal Bjorken flow in three different microscopic theories. Note that for Boltzmann and AdS/CFT, the numerical solutions shown are low dimensional projections from an infinite dimensional space of initial conditions. See text for details.}
\end{figure*}

To elucidate the Borel resummation of the divergent hydrodynamic series it is useful to consider concrete examples. Specifically, let us consider conformal systems undergoing Bjorken flow \cite{Bjorken:1982qr}. Thus, let us assume the system to be homogeneous and isotropic in the coordinates $\tau=\sqrt{t^2-z^2}$ and $\xi={\rm arctanh}\frac{z}{t}$, such that the energy density will only depend on proper time $\tau$. 
As a warm-up, let us review the mock microscopic theory of rBRSSS
\footnote{The acronym BRSSS refers to the conformal hydrodynamic gradient series complete to second order \cite{Baier:2007ix} while rBRSSS denotes the resummed version of BRSSS  \cite{Keegan:2015avk}. This resummation is similar to the one in Muller-Israel-Stewart theory \cite{Muller:1967zza,Israel:1979wp}, but unrelated to the Borel resummation discussed in this work. It is a mock microscopic theory in the sense that results for e.g. the energy density are well behaved even at large gradient strength, yet rBRSSS does not correspond to any real microscopic dynamics.}
(a modern variant of \cite{Muller:1967zza,Israel:1979wp}), where the evolution equation for the energy density is given by $\tau \partial_\tau \ln \epsilon=-\frac{4}{3}+\frac{\Phi}{\epsilon}$, 
$  \tau_\pi \partial_\tau \Phi = \frac{4 \eta}{3 \tau}-\Phi-\frac{4}{3}\frac{\tau_\pi}{\tau} \Phi-\frac{\lambda_1}{2 \eta^2}\Phi^2$ where $\Phi$ is an auxiliary field \cite{Muronga:2001zk,Baier:2007ix,Heller:2015dha,Keegan:2015avk}. Introducing $T(\tau)\propto \epsilon^{1/4}(\tau)$, which customarily has the interpretation of an out-of-equilibrium temperature, it is convenient to consider the dimensionless combinations $C_\eta=\frac{3\eta T}{4\epsilon}$, $C_\pi=\tau_\pi T$ and $C_\lambda=\frac{3\lambda_1 T^2}{4\epsilon C_\eta C_\pi}$ for the three parameters  of rBRSSS. These parameters 
are time-independent because of conformal symmetry. Defining the normalized gradient strength from Eq.~(\ref{eq:NS}) as $\frac{3 \sigma^{\mu\nu}\sigma_{\mu\nu}}{8 T \nabla_\lambda u^\lambda}=\frac{1}{\tau T}$, the rBRSSS equations of motion can be expanded for small gradients $\left(\tau T\right)^{-1}\ll 1$, finding
\begin{equation}
\label{eq:bjorky}
\tau \partial_\tau \ln \epsilon = -\frac{4}{3}+\frac{16 C_\eta}{9 \tau T}+\frac{32 C_\eta C_\pi\left(1-C_\lambda\right)}{27 \tau^2 T^2}+\ldots\,,
\end{equation}
corresponding to the zeroth, first and second-order hydrodynamic gradient series approximation, respectively. In Ref.~\cite{Heller:2015dha}, this gradient series has been extended to $\tau \partial_\tau \ln \epsilon=\sum_{n=0}^{200}\alpha_n \left(\tau T\right)^{-n}$, indeed exhibiting factorial growth for the  coefficients $\alpha_n$. However, the Borel transform ${\cal B}[\tau \partial_\tau \ln \epsilon](\tau T)\equiv \sum_{n=0}^{200}\frac{\alpha_n}{n!} \left(\tau T\right)^{-n}$ exists within a finite radius of convergence around $\left(\tau T\right)^{-1}=0$. ${\cal B}[\tau \partial_\tau \ln \epsilon]$ may be analytically continued to the whole $\tau T$ complex plane by considering the symmetric Pad\'e approximant to ${\cal B}$, finding a dense series of poles with the pole closest to the origin located at $\tau T=z_0^{-1}$, where $z_0=\frac{3}{2 C_\pi}$  \cite{Heller:2015dha}. It is nevertheless possible to define a generalized Borel transform
$
{\cal T}[\tau \partial_\tau \ln \epsilon](\tau T)\equiv \int_{\cal C} dz e^{-z} {\cal B}[\tau \partial_\tau \ln \epsilon](\tau T z)
$
where the contour ${\cal C}$ starts at the origin and ends at $z=\infty$. The ambiguity in the choice of contour in the presence of the singularities of ${\cal B}$ implies an ambiguity in ${\cal T}$, with the pole closest to the origin giving a contribution $\propto e^{-z_0 \tau T}$ to ${\cal T}$. This contribution is non-analytic in $\frac{1}{\tau T}$ and thus responsible for the divergence in the hydrodynamic gradient series, and can be attributed to the presence of a non-hydrodynamic mode. Note that it precisely matches the structure $e^{-\int \frac{d\tau}{\tau_\pi}}$ expected from the known non-hydrodynamic mode in rBRSSS \cite{Baier:2007ix} when using $\tau_\pi^{-1}=\frac{T}{C_\pi}\propto \frac{\tau^{1/3}}{C_\pi}$. The ambiguity in ${\cal T}$ can be resolved by promoting the gradient series to a transseries, e.g. $\tau \partial_\tau \ln \epsilon=\sum_{n,m=0}^\infty c^m \Omega^m(\tau T)\alpha_{nm} \left(\tau T\right)^{-n}$, with $\Omega(\tau T)=(\tau T)^\gamma e^{-z_0\tau T}$, and finding the constants $c,\gamma$ such that the ambiguity in the Borel transform of the transseries part with $m=m_0$ is exactly canceled by $\Omega^{m_0+1}(\tau T)$ for the part with $m=m_0+1$. This program has successfully been performed for rBRSSS in Ref.~\cite{Heller:2015dha,Aniceto:2015mto}. The final result for the Borel transform of $\tau \partial_\tau \ln \epsilon$ can be written in the form $\tau \partial_\tau \ln \epsilon = \left(\tau \partial_\tau \ln \epsilon\right)_{\rm att}+\left(\tau \partial_\tau \ln \epsilon\right)_{\rm non-hydro}$, consisting of a  non-analytic ``attractor'' solution defined for arbitrary $\tau T$ to which the non-hydrodynamic part decays to on a timescale $\tau T\simeq z_0^{-1}$.

Note that obtaining non-analytic solutions from divergent perturbative series' has recently generated considerable interest under the name of ``resurgence'' \cite{Heller:2015dha,Basar:2015ava,Aniceto:2015mto}.

\paragraph{\bf Finding Hydrodynamic Attractors}

Identifying the hydrodynamic attractor solution from the Borel resummation program of the hydrodynamic gradient series is possible, but somewhat tedious. Fortunately, it is possible to obtain the same attractor solution more directly from the equations of motion via the analogue of a slow-roll approximation, cf. Refs.~\cite{Liddle:1994dx,Heller:2015dha} (see Supplemental Material for details). In Fig.~\ref{fig:mis}, results from solving the rBRSSS equations of motions for a range of initial conditions (``numerical'') are as shown together with zeroth, first and second order hydrodynamic gradient series results from Eq.~(\ref{eq:bjorky}). It can be observed that the numerical solutions converge to the hydrodynamic results for moderate gradient strength. One also observes from Fig.~\ref{fig:mis} that the numerical results trend to the unique attractor solution even before matching the gradient series results. This attractor solution is nothing else but the result of the Borel transformation of the divergent transseries as reported in Ref.~\cite{Heller:2015dha}.

\begin{figure*}[t]
  \includegraphics[width=0.5\linewidth]{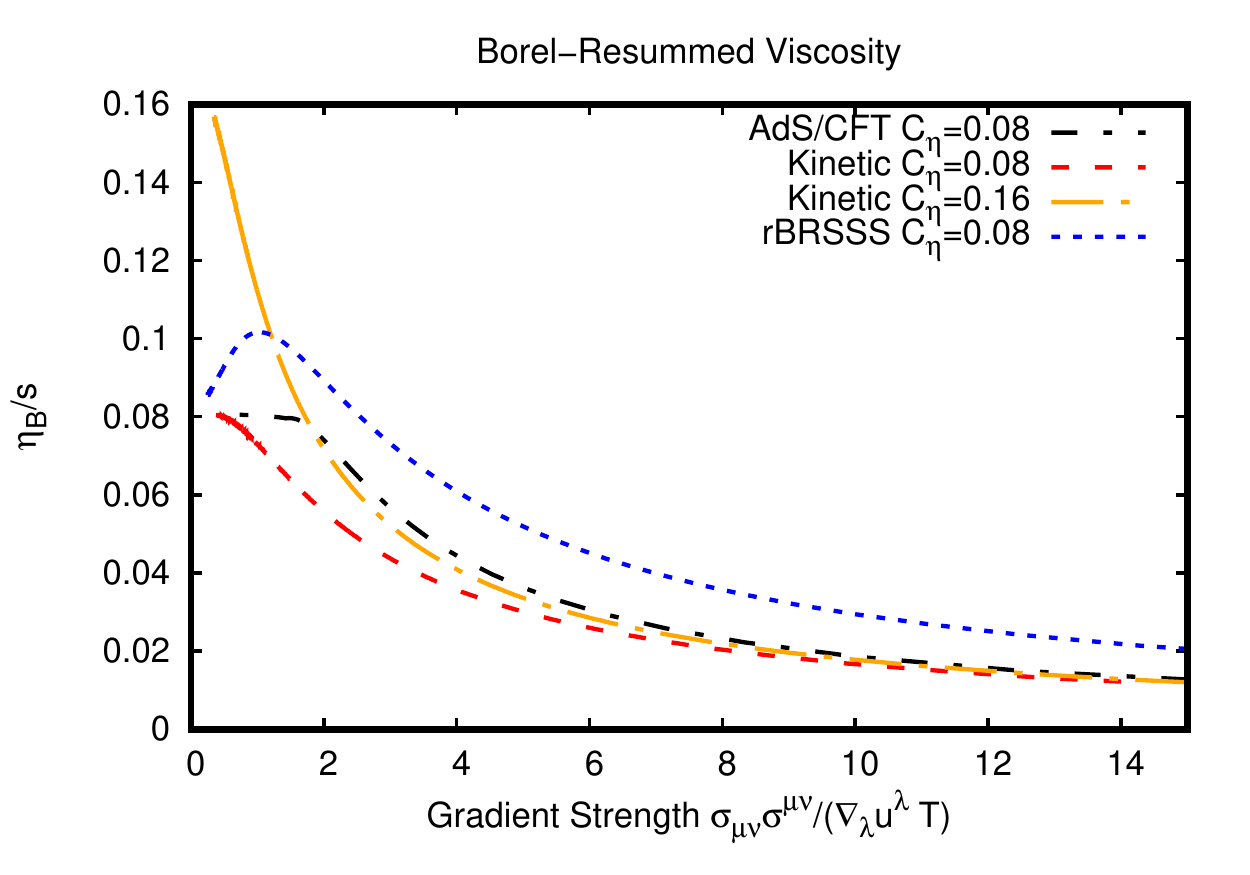}
  \caption{\label{fig:two} Effective viscosity $\eta_B$ versus gradient strength
$\frac{ \sigma^{\mu\nu}\sigma_{\mu\nu}}{ T \nabla_\lambda u^\lambda}$ for attractor solutions in different theories (rBRSSS, kinetic theory and AdS/CFT). For small gradients, one recovers $\frac{\eta_B}{s}=C_\eta$, while $\frac{\eta_B}{s}\rightarrow 0$ for large gradient strength.}
\end{figure*}

\paragraph{\bf Hydrodynamic Attractor in Kinetic Theory}

It is tempting to look for hydrodynamic attractors in other microscopic theories, such as kinetic theory in the relaxation time approximation. This theory is defined by a single particle distribution function $f(t,{\bf x},{\bf p})$ obeying
\begin{equation}
  \label{eq:kinetic}
  p^\mu \partial_\mu f-\Gamma^\lambda_{\mu \nu} p^\mu p^\nu \frac{\partial}{\partial p^\lambda} f=-\frac{f-f^{\rm eq}}{\tau_\pi}\,,
\end{equation}
where here $\Gamma^{\lambda}_{\mu \nu}$ are the Christoffel symbols associated with the Bjorken flow geometry and the equilibrium distribution function may be taken to be $f^{\rm eq}=e^{p^\mu u_\mu/T}$. Here $u^\mu$ is again the time-like eigenvector of $\langle T^{\mu\nu}\rangle=\int \frac{d^3p}{(2\pi)^3} \frac{p^\mu p^\nu}{p} f(x,p)$ and $T$ is the non-equilibrium temperature defined from the time-like eigenvalue of $\langle T^{\mu\nu}\rangle$, which for a single massless Boltzmann particle is $T=\left(\frac{\pi^2 \epsilon}{6}\right)^{1/4}$. 
Note that for a conformal system one can again write $\tau_\pi = C_\pi T^{-1}$ with $C_\pi$ a constant. Solving Eq.~(\ref{eq:kinetic}) numerically, representative results for $\tau \partial_\tau \ln \epsilon$ are shown in Fig.~\ref{fig:mis} (note that $\tau\partial_\tau \ln \epsilon\leq -1$ because the effective longitudinal pressure $P_L=\epsilon \left(1+\tau\partial_\tau \ln \epsilon\right)$ in kinetic theory can never be negative for $f>0$).
%

One observes the same basic structure as in rBRSSS, indicating the presence of a hydrodynamic attractor at early times that arbitrary initial conditions approach via non-hydrodynamic mode decay. (Note that for kinetic theory, the non-hydrodynamic mode is a branch cut giving rise to a decay of the form $e^{-\int \frac{d\tau}{\tau_\pi}}$ \cite{Romatschke:2015gic}).  The attractor solution may be found by finding the initial condition corresponding to a slow-roll approximation at early times and using the numerical scheme to follow the attractor (see Supplemental Material including Refs.~\cite{Baym:1984np,Florkowski:2013lza} for details).
 I find that the kinetic attractor can be approximated by
\begin{equation}
  \label{eq:kinatt}
  \left.\frac{\partial \ln \epsilon}{\partial \ln \tau}\right|^{\rm att}_{\rm kinetic}\simeq - \frac{C_\pi^2+0.744 C_\pi (\tau T) + 0.21 (\tau T)^2}{C_\pi^2+0.6 C_\pi (\tau T) + 0.1575  (\tau T)^2}\,,
\end{equation}
and it coincides with the hydrodynamic solution (\ref{eq:bjorky}) for late times when using the known results $C_\pi=5 C_\eta, C_\lambda=\frac{5}{7}$ for kinetic theory\cite{Romatschke:2011qp,Jaiswal:2014isa}.

\paragraph{\bf Hydrodynamic Attractor in ${\cal N}=4$ SYM} 

Bjorken-flow may also easily be set up in strongly coupled ${\cal N}=4$ SYM in the large number of color limit through the AdS/CFT correspondence. Einstein equations in asymptotic five-dimensional AdS space-time
  $R_{ab}-\frac{1}{2}g_{ab} R-6 g_{ab}=0$
may be solved numerically by the method pioneered by Chesler and Yaffe \cite{Chesler:2008hg},
and the  ${\cal N}=4$ SYM energy-momentum tensor expectation value $\langle T^{\mu\nu}\rangle$ at the conformal boundary can be extracted.  Using the numerical scheme described in Ref.~\cite{Wu:2011yd} (see Supplemental Material for details), results for $\tau \partial_\tau \ln \epsilon$ are shown in Fig.~\ref{fig:mis} compared to the hydrodynamic solutions (\ref{eq:bjorky}) with $C_\eta=\frac{1}{4\pi},C_\pi=\frac{2-\ln 2}{2 \pi},C_\lambda=\frac{1}{2-\ln 2}$ for ${\cal N}=4$ SYM\cite{Baier:2007ix}. Again, the numerical solutions suggest the presence of a hydrodynamic attractor at early times, which is slightly more difficult to see than in the cases of rBRSSS and kinetic theory because the non-hydrodynamic modes for ${\cal N}=4$ SYM are known to have oscillatory behavior (non-vanishing real parts of the black hole quasinormal modes \cite{Berti:2009kk}). Nevertheless, one can discern a preferred attractor candidate without any apparent oscillatory behavior starting at $\lim_{\tau\rightarrow 0}\frac {\partial \ln \epsilon}{\partial \ln \tau}\rightarrow -1$ to which all other initial conditions decay to \footnote{The behavior $\epsilon\propto \frac{1}{\tau}$ for $\tau\rightarrow 0$ seems to contradict the results from Ref.~\cite{Beuf:2009cx}, where it was found that a regular bulk geometry excluded singular behavior of $\epsilon$ for $\tau\rightarrow 0$ if $\epsilon$ possesses a power-series expansion around $\tau=0$. It is possible that the numerical solutions presented here are not sensitive to bulk singularities at $\tau=0$ because the numerics are started at $\tau>0$. Another possibility is that the attractor solution does admit a simple power series expansion for $\epsilon$ around $\tau=0$. Future work is needed to bring clarity to this issue. }. I do not have any analytic understanding of the nature of this AdS/CFT attractor solution, but it is curious to note that it is numerically close to (but clearly different from) the kinetic theory attractor (\ref{eq:kinatt}) with $C_\pi=\frac{5}{4\pi}$. 

\paragraph{\bf Effective Viscosity}

It is possible to interpret the attractor solutions in terms of an effective viscosity coefficient coefficient $\eta_B$ by writing
down a generalized hydrodynamic energy-momentum tensor
\begin{equation}
\label{eq:ffh}
  T^{\mu\nu}_{\rm hydro}=(\epsilon+P_B)u^\mu u^\nu+P_B g^{\mu\nu}-\eta_B \sigma^{\mu\nu}\,
\end{equation}
where for a conformal system $P_B=\epsilon/3$ and $\frac{\eta_B}{s}=\frac{\eta_B}{s}\left(\nabla_\mu\right)$ depends on the local gradient-strength. In the above expressions the subscript 'B' was chosen to indicate Borel-resummed out-of-equilibrium quantities. 
Energy-momentum conservation $u_\nu \nabla_\mu T^{\mu\nu}=0$ leads to $\partial_\tau \ln \epsilon = -\frac{4}{3}+\frac{16 C_\eta}{9 \tau T} \frac{\eta_B}{\eta}$ which can be matched to the hydrodynamic attractor solution, e.g. Eq.~(\ref{eq:kinatt}) to define $\eta_B$ as a function of gradient strength. For the hydrodynamic attractor solutions discussed above, one finds results shown in Fig.~\ref{fig:two}. For small gradients, one recovers $\frac{\eta_B}{s}=\frac{\eta}{s}$, as expected. However, $\eta_B$ eventually tends to zero for far-from-equilibrium systems. 
This finding implies that the effective viscosity $\eta_B$ encountered by an out-of-equilibrium system can be significantly smaller than the equilibrium viscosity $\eta$ calculated from e.g. Kubo relations. Note that this definition of $\eta_B$ is qualitatively similar to Ref.~\cite{Lublinsky:2007mm,Bu:2014ena}, but differs by containing non-linear, but no non-hydrodynamic mode contributions.

\paragraph{\bf Discussion and Conclusions}

In this work, a generalization of fluid dynamics to systems far from local equilibrium was discussed. This generalization rests on the existence of special attractors which become the well-known hydrodynamic solutions once the system comes close to equilibrium. These attractors were explicitly constructed for conformal Bjorken flow for three microscopic theories: rBRSSS (following earlier work in Ref.~\cite{Heller:2015dha}), and for the first time for kinetic theory and strongly coupled ${\cal N}=4$ SYM. For all three systems, it was shown that for arbitrary initial data, attractor solutions are approached via non-hydrodynamic mode decay, demonstrating that the attractor concept is not limited to rBRSSS studied in Ref.~\cite{Heller:2015dha} but applies to a broader class of phenomenologically relevant theories. For conformal systems, attractors can be characterized by an equilibrium equation of state and non-equilibrium viscosity $\eta_B$. Far-from-equilibrium fluid dynamics, defined through Eq.~(\ref{eq:ffh}), constitutes a self-contained set of equations which may prove useful in the description of a broad class of out-of-equilibrium systems. Also, the existence of the attractor solutions for kinetic theory and AdS/CFT suggests the possibility of far-from-equilibrium attractors for the particle distribution function and space-time geometry, respectively.

Many questions remain. Do all microscopic theories possess far-from-equilibrium attractor solutions?   Are attractor functions $\eta_B$ universal for all dynamics of a given microscopic theory? Do non-conformal systems also possess attractors which are characterized by an equilibrium equation of state?  Answering these questions will be the subject of future work.

\begin{acknowledgments}
\section{Acknowledgments}
 
This work was supported in part by the Department of Energy, DOE award No DE-SC0008132. I would like to thank M.~Heller, M.~Spali{\'n}ski and M.~Strickland for fruitful discussions, the organizers of the YITP workshop for Holography, String Theory and Quantum Black Holes for their hospitality and especially W.~Zajc for carefully reading and correcting the manuscript and making many useful suggestions.

\end{acknowledgments}

\bibliography{Borel}

\newpage
\onecolumngrid
\section*{Supplemental Material}

In this supplemental material, details on obtaining the hydrodynamic attractor solutions for rBRSSS, kinetic theory and strongly coupled ${\cal N}=4$ SYM are given.

For rBRSSS, the equations of motion in the case of conformal Bjorken flow $\tau \partial_\tau \ln \epsilon=-\frac{4}{3}+\frac{\Phi}{\epsilon}$, $\tau_\pi \partial_\tau \Phi = \frac{4 \eta}{3 \tau}-\Phi-\frac{4}{3}\frac{\tau_\pi}{\tau} \Phi-\frac{\lambda_1}{2 \eta^2}\Phi^2$ may be decoupled as \cite{Heller:2015dha}
\begin{equation}
\label{eq:fullMIS}
C_\pi \tau T f(\tau T) f^\prime (\tau T)+4 C_\pi C_x(\tau) f^2(\tau T) +\left(\tau T-\frac{16 C_\pi}{3}C_x(\tau) \right)f(\tau T)  
  -\frac{4 C_\eta}{9}+\frac{16 C_\pi}{9}C_x(\tau)
  -\frac{2 \tau T}{3}=0\,,
\end{equation}
where $f(\tau T)\equiv 1+\frac{\tau}{4}\partial_\tau \ln \epsilon$, and $C_x(\tau)\equiv \left(1+\frac{3 C_\lambda \tau T}{8 C_\eta}\right)$ have been introduced for convenience.
Neglecting $f'(\tau T)$ in Eq.~(\ref{eq:fullMIS}) leads to the first approximate attractor solution
\begin{equation}
  \label{eq:misatt1}
  f^{(1)}
\simeq\frac{2}{3}-\frac{\tau T}{8 C_\pi C_x(\tau)}+\frac{\sqrt{64 C_\eta C_\pi C_x(\tau)+9 (\tau T)^2}}{24  C_\pi C_x(\tau)}\,.
\end{equation}
This solution may be consistently improved by iteration. For instance, posing $f^{(2)}=f^{(1)}+\delta f$ and neglecting $\delta f^{\prime}$ in Eq.~(\ref{eq:fullMIS}) leads to a second approximation $f^{(2)}$ for the attractor and so on. In practice, I find the process to converge rapidly such that $f^{(3)}$ is typically a sufficiently accurate approximation to the exact attractor solution for most applications.

For kinetic theory, one would like to find attractor solutions for the energy density. Following seminal work by Baym, Florkowski, Ryblewski and Strickland  \cite{Baym:1984np,Florkowski:2013lza}, it is possible to write down an integral equation for $\epsilon(\tau)$ from the kinetic equations, which takes the form 
\begin{equation}
  \label{eq:kined}
  \epsilon(\tau)=\Lambda_0^4 D(\tau,\tau_0)R\left(\xi(\tau)\right) + \int_{\tau_0}^\tau \frac{d\tau^\prime}{\tau_\pi} D(\tau,\tau^\prime) \epsilon(\tau^\prime)R\left(\frac{\tau^2}{\tau^{\prime 2}}-1\right)\,,
\end{equation}
where $D(\tau_2,\tau_1)=e^{-\int_{\tau_1}^{\tau_2} d\tau^\prime \tau_\pi^{-1}(\tau^\prime)}$, $\xi(\tau)=\left(1+\xi_0\right)\frac{\tau^2}{\tau_0^2}-1$, $R(z)=\frac{1}{2}\left(\frac{1}{1+z}+\frac{\arctan(\sqrt{z})}{\sqrt{z}}\right)$ and $\Lambda_0$ is a typical energy scale. Initial conditions are characterized by a value of $\xi_0$ at $\tau=\tau_0$. Numerical solutions to Eq.~(\ref{eq:kined}) may be obtained by inserting a trial solution $\epsilon^{(0)}(\tau)$ to evaluate the rhs of Eq.~(\ref{eq:kined}), thus finding an improved solution $\epsilon^{(1)}(\tau)$ from the lhs of Eq.~(\ref{eq:kined}), and converging to the exact solution by iterating this process. It is possible to identify points close to  the attractor solution by calculating the equivalent of $f^\prime(\tau T)$ from (\ref{eq:kined}) as
\begin{equation}
  \left.f^\prime\right|_{\tau=\tau_0}\propto\left.\frac{\epsilon\partial_\tau\epsilon +\tau \epsilon \partial_\tau^2\epsilon-\tau \left(\partial_\tau\epsilon\right)^{2}}{\epsilon^2}\right|_{\tau=\tau_0}\,,
  \end{equation}
which does not involve any integrals because the rhs is being evaluated at $\tau=\tau_0$ where e.g. $\epsilon(\tau_0)=R(\xi_0)$. Solving $\left.f^\prime\right|_{\tau=\tau_0}=0$ to obtain
a value of $\xi_0$ at $\tau=\tau_0$ implies that $\epsilon(\tau_0,\xi_0(\tau_0))$ is close to the attractor. One finds for instance $\xi_0(\tau_0=0.1)\simeq 24.5$ for $C_\pi=0.4$. The attractor may then be found from a numerical solution of (\ref{eq:kined}) when using $\xi_0(\tau_0)$ as initial condition at early time $\tau_0\ll 1$.

Let me now give details on how to obtain the attractor solution in strongly coupled ${\cal N}=4$ SYM. Using an ansatz for the five-dimensional line-element
\begin{equation}
\label{eq:ansatz}
  ds^2=2 dr d\tau-A d\tau^2+\Sigma^2 e^{B}dx_\perp^2+\Sigma^2 e^{-2 B}d\xi^2\,
\end{equation}
with $A=A(\tau,r),B=B(\tau,r),\Sigma=\Sigma(\tau,r)$ and $r$ the coordinate in the fifth dimension, it is possible to impose Bjorken-flow at the 4-dimensional Minkowski boundary located at $r\rightarrow \infty$ through $A\rightarrow r^2$, $B\rightarrow -\frac{2}{3}\ln \frac{1+ r\tau}{r}$, $\Sigma^3\rightarrow r^2 (1+r \tau)$. Modifications of these relations at finite $r$ at fixed time correspond to different initial conditions for the time-evolution of $\langle T^{\mu\nu}\rangle$ on the boundary. In particular, the energy density $\epsilon(\tau)\propto-a_4$ is given through the near-boundary series expansion coefficient $a_4(\tau)$ from $A(\tau,r\rightarrow \infty)\simeq r^2\left(1+\sum_{n=1}^\infty a_n r^{-n}\right)$. Initial conditions are specified similar to those in Ref.~\cite{Wu:2011yd}, by choosing $\Sigma^3=r^2\left(1+r \tau\right)+\frac{r^2 s_2+r s_3 +s_4}{r^4+\bar c}$ with $\bar c$ a constant and
\begin{equation}
  s_2(\tau)=\frac{4 \epsilon(\tau)+3 \tau \epsilon^\prime(\tau)}{20}\,,\quad
  s_3(\tau)=\frac{13 \epsilon^\prime(\tau)+5 \tau \epsilon^{\prime\prime}(\tau)}{40}\,,\quad
 s_4(\tau)=\frac{9 \epsilon^\prime(\tau)+135 \tau \epsilon^{\prime\prime}(\tau)+35 \tau^2 \epsilon^{\prime\prime\prime}(\tau)}{560 \tau}
  \end{equation}
at $\tau=\tau_0$. Using the numerical scheme described in Ref.~\cite{Wu:2011yd} and choosing specific initial conditions $\epsilon(\tau)\propto \tau^\alpha$ with constants $\alpha,\bar c$, solving Einstein equations with (\ref{eq:ansatz}) gives numerical results for
$\tau \partial_\tau \ln \epsilon$. Similar to kinetic theory, one can scan initial data parametrized by values of $\bar c,\alpha$ at $\tau=\tau_0$ for numerical results exhibiting weak transients (small $f^\prime$ for all times). One such initial condition is $\alpha\simeq-1,\bar c\simeq 750$ at $\tau_0=0.25$, suggesting that it lies 
close to the hydrodynamic attractor. Once a point close to the attractor has been identified, the attractor solution at later times is again obtained numerically 
using the numerical scheme from Ref.~\cite{Wu:2011yd}.

For convenience, the relevant numerical material used to obtain the attractors in rBRSSS, kinetic theory and AdS/CFT has been made publicly available \cite{codedown}.

\end{document}